\documentclass[prd,aps,twocolumn,tightenlines,notitlepage,nofootinbib,preprintnumbers,letterpaper]{revtex4-1}
\usepackage{epsfig}
\usepackage{amsmath,amsfonts,amssymb,mathrsfs,graphicx,color} 
\usepackage{bm}
\usepackage{hyperref}
\usepackage{graphicx}
\usepackage{amsfonts,amsmath,amssymb,bm}
\usepackage{array}
\usepackage{color}
\usepackage{enumitem}
\usepackage{ulem}
\usepackage[explicit]{titlesec}
\usepackage[usenames,dvipsnames,table]{xcolor}
\usepackage[utf8]{inputenc}
\usepackage{comment}
\usepackage{changepage}
\usepackage[caption=false]{subfig}
\usepackage{subfiles}

\usepackage{accents}

\definecolor{palatd}{RGB}{104, 36, 109}
\definecolor{palatb}{RGB}{0, 56, 168}
\definecolor{palatr}{rgb}{0.745,0.118,0.176}
\newcommand\myshade{80}
\colorlet{mylinkcolor}{palatr}
\colorlet{mycitecolor}{palatb}
\colorlet{myurlcolor}{palatd}

\hypersetup{
  linkcolor  = mylinkcolor!\myshade!black,
  citecolor  = mycitecolor!\myshade!black,
  urlcolor   = myurlcolor!\myshade!black,
  colorlinks = true
}

\begin{document}

\preprint{IFT-UAM/CSIC-25-109, IPPP/25/66}

\setcounter{secnumdepth}{10} 
\title{Updated Constraints on Large Extra Dimensions from Reactor Antineutrino Experiments}

\author{T. Gökalp Elaçmaz}%
\email{gokalp.elacmaz@bilkent.edu.tr}
\affiliation{Department of Physics, Bilkent University, 
Bilkent 06800, Ankara, Türkiye}

\author{Ivan Martinez-Soler}%
\email{ivan.j.martinez-soler@durham.ac.uk}
\affiliation{Institute for Particle Physics Phenomenology, Durham University, South Road DH1 3LE, Durham, U.K.}

\author{Yuber F. Perez-Gonzalez}%
\email{yuber.perez@uam.es}
\affiliation{Departamento de Física Teórica and Instituto de Física Teórica UAM-CSIC, Universidad Autónoma de Madrid, Cantoblanco, 28049 Madrid, Spain}

\begin{abstract}
    We investigate constraints on large extra dimensions (LED) using the latest results from reactor antineutrino experiments. 
    Specifically, we analyze the full data sets from Daya Bay, RENO, KamLAND, NEOS, and STEREO to derive updated bounds. 
    For the case of one extra dimension, we find constrains on its radius $a$ of $a \lesssim 0.58~{\rm \mu m}$ ($a \lesssim 0.12~{\rm \mu m}$) at the $99\%$ confidence level for normal (inverted) ordering, an improvement of approximately $\sim 20\%$ ($\sim 25\%$) with respect to previous bounds, assuming a massless lightest active neutrino.
    Furthermore, we present new limits on $4+d$ LED scenarios, with $d = 2, 3, 4$ denoting the number of extra dimensions, based on the same reactor data and assuming equal radii for all extra dimensions. 
    We find that the constraints become increasingly stringent with a larger number of extra dimensions. 
    In particular, $d = 4$ with a massless lightest active neutrino, we obtain limits of  $a \lesssim 0.28~{\rm \mu m}$ for normal and  $a \lesssim 0.05~{\rm \mu m}$ for inverted orderings at the $99\%$ confidence level.
\end{abstract}

\maketitle

\section{Introduction}\label{sec:intro} 

The experimental confirmation that neutrinos possess mass, established through the observation of neutrino oscillations~\cite{Super-Kamiokande:1998kpq,SNO:2002tuh}, has motivated efforts to understand the origin of the mass hierarchy between charged leptons and neutrinos. 
The most prominent mechanism proposed to explain this disparity is the \textit{see-saw} model~\cite{Mohapatra:1979ia,Gell-Mann:1979vob,Yanagida:1979as,Minkowski:1977sc,Mohapatra:1980yp,Magg:1980ut,Lazarides:1980nt,Wetterich:1981bx,Foot:1988aq,Ma:1998dn}, which introduces right-handed neutrinos, singlet states under the Standard Model (SM) gauge group, into the particle spectrum. 
Because these singlet fields are not constrained by SM gauge symmetries, they can acquire Majorana mass terms whose scale is not protected and may therefore take arbitrary values. 
If these Majorana masses are of the order of a grand unified theory (GUT) scale, ${\cal O}(10^{12}~{\rm GeV})$, the resulting left-handed neutrino masses are naturally suppressed relative to the electroweak scale, thereby accounting for their observed smallness.
In this scenario, it is therefore predicted that neutrinos are of Majorana nature.

There are, however, alternative scenarios that can account for the smallness of neutrino masses. 
An interesting possibility is the existence of additional compactified spatial dimensions.
This idea originally emerged as an attempt to address the SM hierarchy problem, that is, the large disparity between the Higgs mass and the Planck scale~\cite{Arkani-Hamed:1998jmv}. 
If gravity is allowed to propagate through the extra dimensions while the SM fields remain confined to a four-dimensional brane, gravity appears weaker than the other fundamental interactions, effectively pushing the observed Planck scale to a large value~\cite{Arkani-Hamed:1998jmv}.

In a similar manner, if the right-handed degrees of freedom are allowed to propagate through the extra dimensions, the active neutrino masses become suppressed with respect to the electroweak scale~\cite{Arkani-Hamed:1998wuz}. 
Since the presence of large extra dimensions (LED) leads to the appearance of a tower of states in the four-dimensional brane, known as Kaluza-Klein (KK) modes, neutrino oscillations can be significantly modified as the KK states mix with active neutrinos~\cite{Arkani-Hamed:1998wuz,Davoudiasl:2002fq,Cao:2003yx}. 
Therefore, depending on the LED parameters, active-sterile oscillations may occur and can be constrained by experimental data.

As such, limits on LED from neutrino data have been derived from a variety of sets, including short- and long-baseline experiments~\cite{Machado:2011jt,Carena:2017qhd}, solar anomalies~\cite{Machado:2011kt,Forero:2022skg}, IceCube measurements~\cite{Esmaili:2014esa}, beta-decay data~\cite{Basto-Gonzalez:2012nel,Rodejohann:2014eka,Antoniadis:2025rck}, cosmological observations~\cite{McKeen:2024fdl,Anchordoqui:2024xvl}, colliders~\cite{Cao:2004tu}, and reactor antineutrino experiments~\cite{Machado:2011jt,Girardi:2014gna,Forero:2022skg,Eller:2025lsh}. 
Also, there exist some forecast for future experiments such as DUNE~\cite{Berryman:2016szd,Siyeon:2024pte}, SNBD~\cite{Stenico:2018jpl}, JUNO~\cite{Basto-Gonzalez:2021aus}, P2SO~\cite{Panda:2024ioo}, or nuSTORM~\cite{Franklin:2025muw}.
Current constraints are dominated by the MINOS and Daya Bay experiments, depending on the neutrino mass ordering~\cite{Forero:2022skg}. 
Moreover, most of the derived bounds have been obtained under the assumption of a single LED. 
This assumption is justified by considering that any additional extra dimensions would have smaller radii, and therefore would not affect the oscillation pattern.

In this work we update the constraints coming from reactor antineutrinos on LED from the latest results of Daya Bay~\cite{DayaBay:2019yxq,DayaBay:2022orm}, RENO~\cite{RENO:2024msr}, KamLAND~\cite{KamLAND:2013rgu}, STEREO~\cite{STEREO:2019ztb}, and NEOS~\cite{RENO:2020hva} experiments.
For a negligible mass of the lightest neutrino, we find that for the normal ordering (NO) the bound is still dominated by MINOS, while for the inverted ordering (IO) the bound is improved by about $30\%$.
Additionally, we go beyond the assumption of the dominance of a single extra dimension following the procedure established in Refs.~\cite{Cao:2003yx,Cao:2004tu} to compute the oscillation probabilities. We derive limits for $4+d$ dimensions, with $d=1,\ldots,4$, assuming all extra dimensions have the same radius.

This paper is organized as follows. In Sec.~\ref{sec:LED}, we describe the LED model in the $4+d$ scenario and the computation of the oscillation probability in such a case.
In Sec.~\ref{sec:sim}, we describe the generalities of the reactor antineutrino simulations used in the determination of our bounds. Finally in Sec.~\ref{sec:results} and Sec.~\ref{sec:concl} we present our results and conclusions, respectively. 
App.~\ref{sec:simdet} contains the detailed description of the simulation of the reactor experiments used in this work.
Throughout this manuscript, we adopt the natural units system in which $\hbar = c = k_{\rm B} = 1$.

\section{Large Extra Dimensions}\label{sec:LED}

Large extra dimensions (LED) have been proposed as a potential solution to the long-standing hierarchy problem~\cite{Arkani-Hamed:1998jmv,Antoniadis:1998ig}. This proposal is based on the assumption that gravity, unlike the Standard Model (SM) fields, is not confined to our familiar 4-dimensional spacetime (or \textit{brane}), but can also propagate through additional spatial dimensions. Meanwhile, SM fields remain restricted to the 4-dimensional brane.
Under this scenario, gravity appears significantly weaker than the other fundamental forces to an observer confined to the brane. This weakness arises because gravity spreads into the extra dimensions, effectively diluting its strength in our 4D spacetime. The observed 4-dimensional Planck scale, $M_4$, is related to the fundamental Planck scale in $(4 + d)$ dimensions, $M_d$, by the relation
\begin{align}
    M_4^2 = M_d^{2 + d} V_d,
\end{align}
where $V_d$ is the volume of the compactified $d$ extra dimensions.

If the fundamental scale is of the order of the TeV scale, $M_d \sim \mathrm{TeV}$, and the extra dimensions are compactified in the form of a $d$-dimensional torus such that $V_d = (2\pi a)^d$, with $a$ denoting the radius of the extra dimensions, then the observed 4-dimensional Planck scale, $M_4 = 2 \times 10^{18}~\mathrm{GeV}$, is recovered if the extra-dimensional volume is large enough. For instance, in the case of a single extra dimension ($d = 1$), this requires a radius of approximately $a \sim 6 \times 10^{13}~\mathrm{cm}$~\cite{Rattazzi:2003ea}.
Such a large value for the extra-dimensional radius is ruled out by observations. However, considering the case of two extra dimensions ($d = 2$), the required radius reduces to approximately $a \sim 0.4~\mathrm{mm}$. This range is already being probed by experiments designed to detect deviations from the gravitational inverse-square law, leading to an experimental upper bound of $a \lesssim 20~\mathrm{\mu m}$. This constraint, in turn, implies a lower bound on the fundamental Planck scale, $M_d \gtrsim 4~\mathrm{TeV}$~\cite{Rattazzi:2003ea}. Since the fundamental Planck scale would be close to the electroweak scale, $M_d \sim G_F^{-1/2}$, where $G_F$ is the Fermi constant, corrections to the Higgs mass would be of the same order as its mass, solving the hierarchy problem.

Beyond addressing the weakness of gravity, LED models have also been proposed to explain the smallness of neutrino masses~\cite{Dienes:1998sb,Arkani-Hamed:1998wuz}. In this class of models, right-handed neutrinos, as SM gauge singlets, are allowed to propagate in the extra dimensions. Their Yukawa interactions with lepton and Higgs doublets then give rise to neutrino masses compatible with observations. These same interactions also induce mixing between active neutrinos and the Kaluza-Klein (KK) modes of the bulk fermions, which appear on the 4-dimensional brane as sterile neutrinos, leading to observable deviations in the oscillation patterns measured in various experiments~\cite{Davoudiasl:2002fq,Cao:2003yx,Cao:2004tu,Machado:2011jt,Esmaili:2014esa,Carena:2017qhd,Forero:2022skg}.

In the following, we adopt the formalism presented in Ref.~\cite{Cao:2003yx}, which provides a general derivation of the oscillation probability in scenarios with $d$ extra dimensions.
We begin with the general action $\mathscr{S}$ defined in the full $(4 + d)$-dimensional spacetime,
\begin{align}
    \mathscr{S} = \int d^4x\, d^d y\, \left[\mathscr{L}_{\rm Bulk} + \delta(y)\, \mathscr{L}_{\rm Brane}\right],
\end{align}
where $\mathscr{L}_{\rm Bulk}$ describes the dynamics of fields propagating in the $(4 + d)$-dimensional bulk, namely gravity and right-handed neutrinos, while $\mathscr{L}_{\rm Brane}$ encodes the interactions of the SM fields confined to the brane, localized at $y = 0$.
For the neutrino sector, the relevant terms are given by
\begin{subequations}
    \begin{align}
        \mathscr{L}_{\rm Bulk} & \supset \bar{\Psi}_\alpha i \Gamma^M D_M \Psi_\alpha,\\
        \mathscr{L}_{\rm Brane} & \supset - \frac{Y_{\alpha \beta}}{M_d^{d/2}} H\, \overline{\nu}_{\alpha L} \psi_{\beta R} + \text{h.c.},
    \end{align}
\end{subequations}%
where $\Psi_\alpha$ denotes three fermionic fields propagating in $(4 + d)$ dimensions, $\Gamma^M$ are the Dirac matrices, and $D_M$ is the covariant derivative, all defined in $(4 + d)$ dimensions. 
On the brane-localized Yukawa interactions contained in $\mathscr{L}_{\rm Brane}$, $Y_{\alpha \beta}$ are the Yukawa couplings, assumed to be of ${\cal O}(1)$, $H$ is the Higgs field, $\overline{\nu}_{\alpha L}$ are the active neutrino fields, and $\psi_{\beta R}$ corresponds to the right-handed, four-dimensional projection of the bulk fields $\Psi_\alpha$.

Performing the KK expansion as described in Ref.~\cite{Cao:2003yx}, we obtain the 4-dimensional neutrino mass Lagrangian:
\begin{align}\label{eq:4dmass_lag}
    \mathscr{L}_{\rm mass}^{(4)} &= \sum_{\alpha=1}^3 \sum_{\hat n} \frac{|\hat n|}{a} \, \overline{\nu}_{\alpha L}^{(\hat n)} \nu_{\alpha R}^{(\hat n)} \notag\\ 
    & + \sum_{\alpha, \beta} m_{\alpha \beta}^D \left[ \overline{\nu}_{\alpha L}^{(0)} \nu_{\beta R}^{(0)} + \sqrt{2} \sum_{\hat n} \overline{\nu}_{\alpha L}^{(0)} \nu_{\beta R}^{(\hat n)} \right] + \text{h.c.},
\end{align}
where the fields $\{\nu_{\alpha R}^{(0)}, \nu_{\alpha R}^{(\hat n)}, \nu_{\alpha L}^{(\hat n)}\}$ are defined as
\begin{subequations}
    \begin{align}
        \nu_{\alpha R}^{(0)} &\equiv \psi_{\alpha R}^{(0)}, \\
        \nu_{\alpha R}^{(\hat n)} &\equiv \frac{1}{\sqrt{2}} \bigl(\psi_{\alpha R}^{(\hat n)} + \psi_{\alpha R}^{(-\hat n)}\bigr),
        \quad \hat n = 1 \dots \infty, \\
        \nu_{\alpha L}^{(n)} &\equiv \frac{1}{\sqrt{2}} \bigl(\psi_{\alpha L}^{(\hat n)} + \psi_{\alpha L}^{(-\hat n)}\bigr),
        \quad \hat n = 1 \dots \infty.
    \end{align}
\end{subequations}
In Eq.\eqref{eq:4dmass_lag}, the Dirac masses $m_{\alpha \beta}^D$ are 
\begin{align}
    m_{\alpha \beta}^D = \frac{Y_{\alpha \beta}\, v}{(V_d M_d^{d})^{1/2}},
\end{align}
where $v$ is the Higgs vacuum expectation value. From this expression, it follows that neutrino masses are suppressed relative to the electroweak scale $v$ if the product $V_d M_d^d$ is large, even when the Yukawa couplings $Y_{\alpha \beta}$ are of order one, offering a natural explanation for the smallness of neutrino masses.

Additionally, $\hat n^2 \equiv \sum_{i=1}^d n_i^2$, with $n_i = 0,1,\ldots,\infty$, where each $n_i$ labels one of the extra dimensions, excluding the combination where all $n_i = 0$ in Eq.~\eqref{eq:4dmass_lag}~\cite{Cao:2003yx}. 
As a result, there are $k_{\hat n}$ degenerate modes with coupling $|\hat n|/a$. For example, in the case of $d = 2$, the states with mass $5/a$ have a multiplicity of $k_{\hat n} = 4$, since the combinations $(n_1, n_2) = \{(0,5), (3,4), (4,3), (5,0)\}$ all yield the same value of $|\hat n|$.

Clearly, the mass Lagrangian is still not diagonal. The mass matrix is 
\begin{align}
    M =
    \begin{pmatrix}
    m_i^D & 0         & \cdots        & 0^{\times k_{\hat{n}}} & \cdots \\
    \sqrt{2}\,m_i^D & \tfrac{1}{a} & \cdots        & \cdots & \cdots \\
     \vdots    & \vdots    & \ddots  & \vdots & \vdots \\
    (\sqrt{2}\,m_i^D)^{\times k_{\hat{n}}\, T} & 0^{\times k_{\hat{n}}\, T}    &  \cdots & \tfrac{|\hat n|}{a} \cdot I_{k_{\hat{n}}} &  \cdots \\
    \vdots    & \vdots  & \vdots & \vdots & \ddots  \\
    \end{pmatrix},
\end{align}
where the notation $x^{\times d}$ denotes the value $x$ repeated $d$ times as entries of a row or column vector, and $i=1,2,3$ is a generation index.
To diagonalize the mass matrix, two unitary matrices, $L_i$ and $R_i$, must be determined such that $R_i^\dagger M_i L_i$ is diagonal. The mass eigenstates are then defined by $\tilde{\nu}_L^{i} = L_i^\dagger \, \nu_L^{'i} = R_i^\dagger \nu_R^{'i}$. Defining $\xi_i = \sqrt{2} m_i a$, and noting that $L_i$ consists of the eigenvectors of $M_i^\dagger M_i$, one can obtain $L_i$. The eigenvalues and eigenvectors of the following matrix will be relevant for this purpose~\cite{Machado:2011jt}
\begin{align}
    X &= a^2 M^\dagger M \notag\\
    & =
    \lim_{N\to\infty}\begin{pmatrix}
    \left(\mathscr{D}_d + \tfrac{1}{2}\right)\xi_i^2 & \xi_i & \dots & (|\hat{n}|\, \xi_i)^{\times k_{\hat{n}}} & \dots & N\xi_i \\[6pt]
    \xi_i & 1 & \dots & \dots & \dots & 0 \\[6pt]
    \vdots & \vdots & \ddots  & \vdots& \vdots   & \vdots \\[6pt]
    \left(|\hat{n}|\, \xi_i\right)^{\times k_{\hat{n}}\, T} & \dots & \dots & |\hat{n}|^2 \cdot I_{k_{\hat{n}}} & \dots & 0 \\[6pt]
    \vdots  & \vdots & \vdots & \vdots & \ddots & \vdots \\[6pt]
    N\xi_i & 0 & \dots  & 0 & \dots & N^2
    \end{pmatrix}.
\end{align}
where $\mathscr{D}_d = \sum_i k_i$, such that for $d=1$, $\mathscr{D}_1 = N$.

As noted in Ref.~\cite{Cao:2003yx}, certain caveats must be considered in the treatment of neutrino oscillations. First, the KK expansion must be truncated at the cutoff scale $M_d$, beyond which the effective theory breaks down. For $d > 1$, the KK sum may also diverge due to the increasing degeneracy of the $\hat{n}_k$ states~\cite{Cao:2003yx}.
Finally, one may ask whether higher-dimensional operators or a UV completion could suppress the off-diagonal mass terms, thereby weakening oscillation constraints. However, since the same operator generates both diagonal and off-diagonal terms, such suppression would conflict with experimental bounds. As a result, the constraints on $1/a$ remain robust for $d \leq 3$, although they may exhibit mild dependence on the UV completion~\cite{Cao:2003yx}.

Once the diagonalization is performed, the transition probability $P(\nu_\alpha^{(0)} \to \nu_\beta^{(0)}; L)$ is given by
\begin{align}\label{eq:prob}
    P(\nu_\alpha^{(0)} \to \nu_\beta^{(0)}; L) = \left| \mathscr{A}(\nu_\alpha^{(0)} \to \nu_\beta^{(0)}; L) \right|^2.
\end{align}
The corresponding transition amplitude for a neutrino with energy $E$ traveling a distance $L$ is given by
\begin{widetext}
\begin{align}
    \mathscr{A}(\nu_\alpha^{(0)} \to \nu_\beta^{(0)}; L) 
    = \sum_{i,k=1}^3 U_{\alpha i}\, U_{\beta k}^* \sum_{j=1}^3 \sum_{n = 0} W_{ij}^{(0 n)\,*} W_{kj}^{(0 n)} \exp\left(i\frac{(\lambda_j^{(n)})^2 L}{2 E a^2}\right),
\end{align}
where $\lambda_j^{(n)}$ are the Hamiltonian eigenvalues while the matrices $U$ and $W$ describe the mixing among active neutrino flavors and between active and KK modes, respectively.
In general, the Hamiltonian in the mass basis takes the form
\begin{align}
    H = \frac{1}{2E} M^\dagger M + 
    \begin{pmatrix}
        V_{3\times 3} & \mathbf{0}_{3\times 3\mathscr{D}_d} \\
        \mathbf{0}_{3\mathscr{D}_d \times 3} & \mathbf{0}_{3\mathscr{D}_d \times 3\mathscr{D}_d}
    \end{pmatrix},
\end{align}
where $V$ accounts for possible matter effects during neutrino propagation and is given by~\cite{Machado:2011jt}
\begin{align}
    V_{ij} = \sum_{\alpha=e,\mu,\tau} U_{\alpha i}^*\, U_{\alpha j} V_{\alpha},
\end{align}
with $V_{\alpha} = \sqrt{2}G_F\left\{\delta_{e\alpha} n_e - n_n/2 \right\}$. Here, $n_e$ and $n_n$ denote the electron and neutron number densities in the medium, respectively. The symbol $\mathbf{0}_{x \times y}$ represents a zero matrix of dimension $x \times y$.

In the limit $\xi_i \ll 1$ and assuming vacuum propagation, approximate expressions for the eigenvalues can be obtained, see Ref.~\cite{Cao:2003yx}, 
\begin{align}
    \lambda_j^{( n)} = \begin{cases}
        \dfrac{\xi_j}{\sqrt{2}} \left(1 - \dfrac{\xi_j^2}{2} \displaystyle\sum_{ n^\prime=1}^\infty \dfrac{k_{n^\prime}}{( n^\prime)^2} \right) & \text{for }  n = 0, \\
        n \left(1 + \dfrac{\xi_j^2}{2 n^2} k_{n} \right) & \text{for } n > 1 \text{ and the 1 state}, \\
         n & \text{for } n > 1 \text{ and the remaining } (k_{n} - 1) \text{ states}.
    \end{cases}
\end{align}
In vacuum, the elements of the mixing matrix take the form $W_{ij}^{(0 n)} = W_{i}^{(0 n)} \delta_{ij}$, indicating that KK modes associated with different generations decouple. This matrix coincides with the matrix $L_i$ defined earlier and is given by
\begin{align}
    W^{(00)}_j = 1 - \sum_{\hat n^\prime=1}^\infty \frac{\xi_j^2}{2(\hat n^\prime)^2} k_{\hat n^\prime}, \qquad 
    W^{(0 n)}_j = \frac{\xi_j}{n}, \quad\text{for}~n>1.
\end{align}
Note that for $d = 1$, we have $k_{n} = 1$ for all $n$, so that $\displaystyle\sum_{n^\prime = 1}^\infty |n^\prime|^{-2} = \pi^2 / 6$, reproducing the same eigenvalues and mixing approximations given in Refs.~\cite{Davoudiasl:2002fq,Machado:2011jt,Esmaili:2014esa,Carena:2017qhd,Forero:2022skg}.
For $d>1$, however, the previous approximations are only valid if the sum $\displaystyle\sum_{n^\prime = 1}^\infty k_{n^\prime}|n^\prime|^{-2}$ converges, which might not be the case.
In the same limit, the oscillation amplitude can be written as
\begin{align}\label{eq:approx_amp}
    \mathscr{A}(\nu_\alpha^{(0)} \to \nu_\beta^{(0)}; L) &= \sum_{i=1}^3 U_{\alpha i}\, U_{\beta i}^* \sum_{n=0}^\infty |W_{i}^{(0n)}|^2\exp\left(i\frac{(\lambda_{i}^{(n)})^2 L}{2 E}\right)\notag\\
    &\approx \sum_{i=1}^3 U_{\alpha i}\, U_{\beta i}^* \exp\left(i\frac{\Delta m_{i1}^2 L}{2 E}\right)+ 2\sum_{i=2,3} U_{\alpha i}\, U_{\beta i}^* \sum_{n = 1}^\infty \frac{\Delta m_{i1}^2 a^2}{n^2} k_n \exp\left(i\frac{n^2 L}{2 E a^2}\right)\notag\\
    &\quad + 2\sum_{i=1}^3 U_{\alpha i}\, U_{\beta i}^* \sum_{n = 1}^\infty \frac{m_1^2 a^2}{n^2} k_n \exp\left(i\frac{n^2 L}{2 E a^2}\right),
\end{align}
\end{widetext}
where $\Delta m_{ij}^2 = m_i^2 - m_j^2$ as usual, and multiplied by an irrelevant phase $\exp(-i\lambda_{1}^{(0)}L/(2Ea^2))$. For clarity, we have taken $W_j^{(00)} \approx 1$, $\lambda_j^{(0)} \approx \xi_j/\sqrt{2}$, and $\lambda_j^{(n)} \approx n$ for all $n > 1$.
As pointed out in Ref.~\cite{Carena:2017qhd}, the first term corresponds to the standard three-neutrino oscillation amplitude, while the remaining terms represent contributions from the extra-dimensional modes. These additional terms introduce interference effects that deviate from the standard oscillation pattern.
The factor $k_n$, which appears in the presence of extra dimensions beyond $d = 1$, enhances the contribution of KK modes to the mixing. As a result, for $d > 1$, larger deviations from standard oscillations are expected due to the increased number of states involved in the propagation. This can also be interpreted as an effective enhancement of the active-sterile mixing angle, leading to a greater disappearance of active neutrinos.

\begin{figure*}[t!]
    \centering
    \includegraphics[width=1\linewidth]{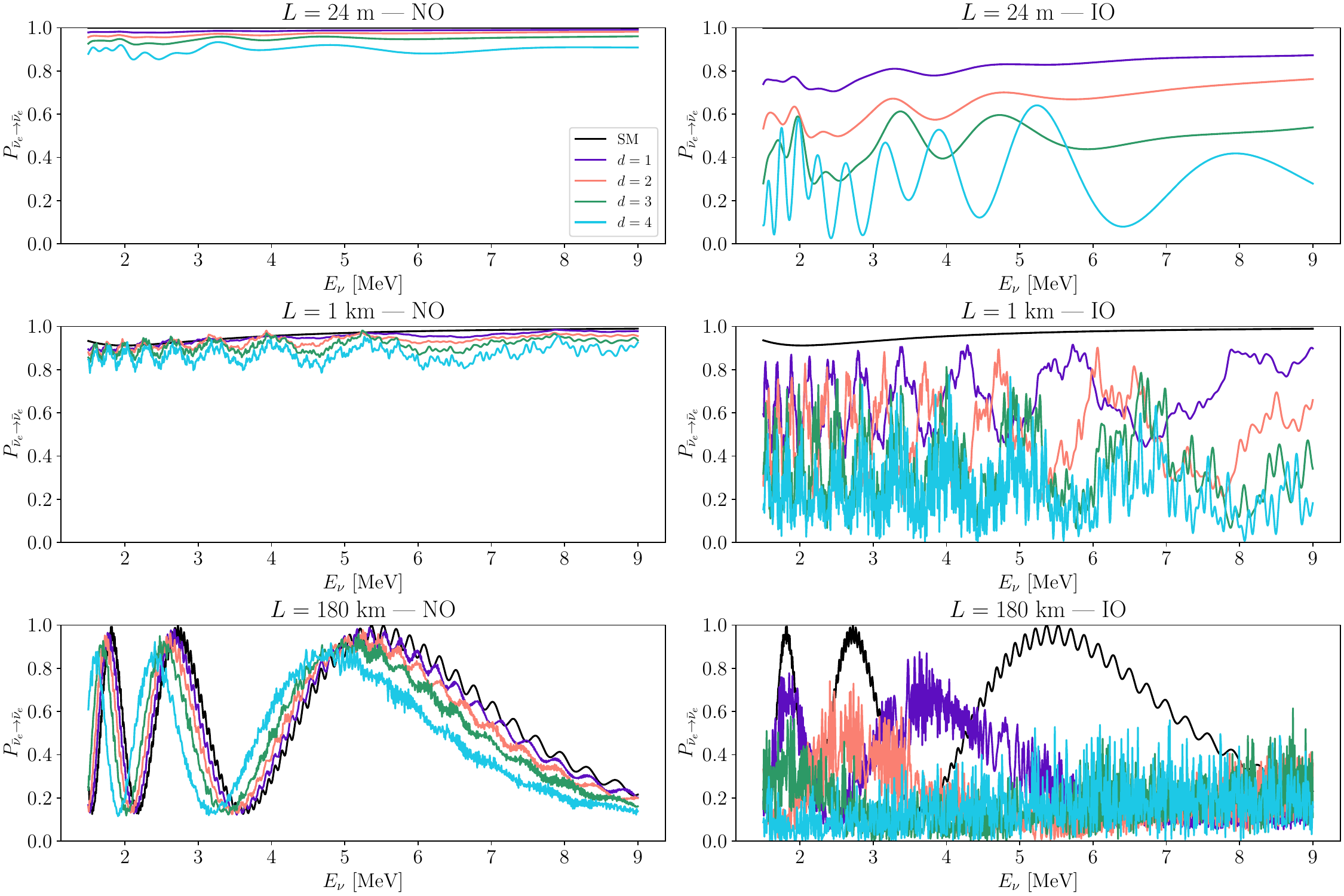}
    \caption{Electron antineutrino survival probability as a function of energy in the presence of LED, shown for $d = 1$ (purple), $d = 2$ (orange), $d = 3$ (green), and $d=4$ (turquoise). Each row corresponds to a different baseline: $L = 24$~m (top), $L = 1$~km (middle), and $L = 180$~km (bottom). Left and right panels show results for normal and inverted mass orderings, respectively. The parameters are fixed to $a = 1~\mu\mathrm{m}$ and $m_0 = 0.01$~eV.}
    \label{fig:probs}
\end{figure*}
In Fig.~\ref{fig:probs}, we show the electron antineutrino survival probability in the presence of LED for different numbers of extra dimensions: $d = 1$ (purple), $d = 2$ (orange), and $d = 3$ (green). We consider three baselines, $L = 24$~m (top), $L = 1$~km (middle), and $L = 180$~km (bottom), for both normal (left panels) and inverted (right panels) mass orderings. The parameters are fixed to $a = 1~\mu\mathrm{m}$ and a lightest neutrino mass of $m_0 = 0.01$~eV.
At short baselines, where standard oscillations are not expected, increasing the number of extra dimensions leads to a suppression of the survival probability. For instance, in the case of $d = 3$ and inverted ordering, the probability can drop to $P(\bar{\nu}_e \to \bar{\nu}_e) \approx 0.25$. In contrast, for normal ordering, the suppression is milder, with the probability remaining above $P(\bar{\nu}_e \to \bar{\nu}_e) \approx 0.9$.
This deviation from the standard three-neutrino expectation becomes more pronounced as the baseline increases. Additionally, we observe oscillatory features arising from the contributions of higher KK modes (i.e., larger values of $n$). At the longest baseline considered, $L = 180$~km, an additional effect appears, a shift in the oscillation minima due to modifications in the Hamiltonian eigenvalues induced by the presence of LED.

We now briefly comment on the difference between normal and inverted ordering. As discussed in Refs.~\cite{Machado:2011jt}, the contribution from KK states to the oscillation amplitude scales as $\mathscr{A}(\nu_\alpha \to \nu_\beta) \propto \sum_i \xi_i^2 U_{\alpha i} U_{\beta i}^*$, meaning that the PMNS matrix elements are weighted by the squared mass of the corresponding neutrino state. In the NO case, the largest mass term $\xi_3$ is suppressed by the small value of $\sin^2 \theta_{13}$, resulting in a less pronounced LED effect. On the other hand, in the IO case, the dominant contributions come from $\xi_1$ and $\xi_2$, which are not suppressed, leading to stronger deviations.

Having analyzed the impact of LED on oscillation probabilities, we now turn to its implications for neutrino phenomenology, focusing specifically on reactor antineutrino experiments.
In the following, we numerically diagonalize the full Hamiltonian to obtain the mass eigenvalues and mixing matrix for $d\geq 1$. The KK tower is truncated at $N = 6$, as including additional modes was found to have a negligible impact on the results.

\section{Reactor Simulations}\label{sec:sim}

In reactor neutrino experiments, a flux of electron antineutrinos is produced with energies in the MeV range. This flux originates from the fission of four isotopes, primarily $^{235}$U and $^{239}$Pu, with additional contributions from $^{238}$U and $^{241}$Pu. Several theoretical calculations of the reactor neutrino flux have been performed. In our case, for the two dominant isotopes, we use spectra extracted from reactor measurements~\cite{DayaBay:2019yxq}, while for the remaining two, we rely on predictions from the Huber-Mueller model.

The reactor neutrino flux has been measured across a wide range of baselines, from $\sim 10$~m~\cite{RENO:2020hva,STEREO:2019ztb} to several hundred kilometers~\cite{KamLAND:2013rgu}. These measurements have enabled the most precise determinations of some oscillation parameters. For instance, kilometer scale measurements have provided precise determination of parameters such as $\sin^2\theta_{13}$ and $\Delta m^2_{31}$~\cite{DayaBay:2022orm,RENO:2024msr}. Longer-baseline experiments have allowed for the determination of $\Delta m^2_{21}$~\cite{KamLAND:2013rgu}. Meanwhile, short-baseline experiments have been used to search for oscillations at shorter wavelengths, induced by eV sterile neutrino~\cite{STEREO:2019ztb,RENO:2020hva,Giunti:2021kab,Hardin:2022muu}. Future measurements of reactor neutrino fluxes will further improve our understanding of neutrino evolution~\cite{JUNO:2022mxj,Parke:2024xre,deGouvea:2020vww}.

For this analysis, we use the full datasets from Daya Bay and RENO at the $1~{\rm km}$ baseline. For long-baseline measurements, we include KamLAND~\cite{KamLAND:2013rgu}, and for short-baseline experiments, we incorporate NEOS and STEREO. Although we use the same reactor flux model across all these analyses, we adjust the experimental efficiencies, exposure times, and systematic uncertainties to accurately reproduce the observed spectral data. See App.~\ref{sec:simdet} for a detailed description of the simulations used.

In all these experiments, neutrinos are detected via inverse beta decay (IBD), $\overline{\nu_{e}} + p \rightarrow n + e^{+}$. The annihilation of the positron gives rise a two photon signal (prompt-signal) with a total energy of $1.022$~MeV. The posterior capture of the neutron by the hydrogen leads to a photon emission of $2.2$~MeV gamma ray, called delayed signal. The prompt energy ($E_{p}$) is related to positron kinetic energy by, $E_{p} = T_{e} + 1.022$~MeV. Assuming the neutron is at rest, the neutrino energy can be inferred from the prompt signal by $E_\nu \simeq E_{p} + 0.78$. In the analysis, the event distribution is presented as a function of the prompt energy.

\begin{figure}
    \centering
    \includegraphics[width=1\linewidth]{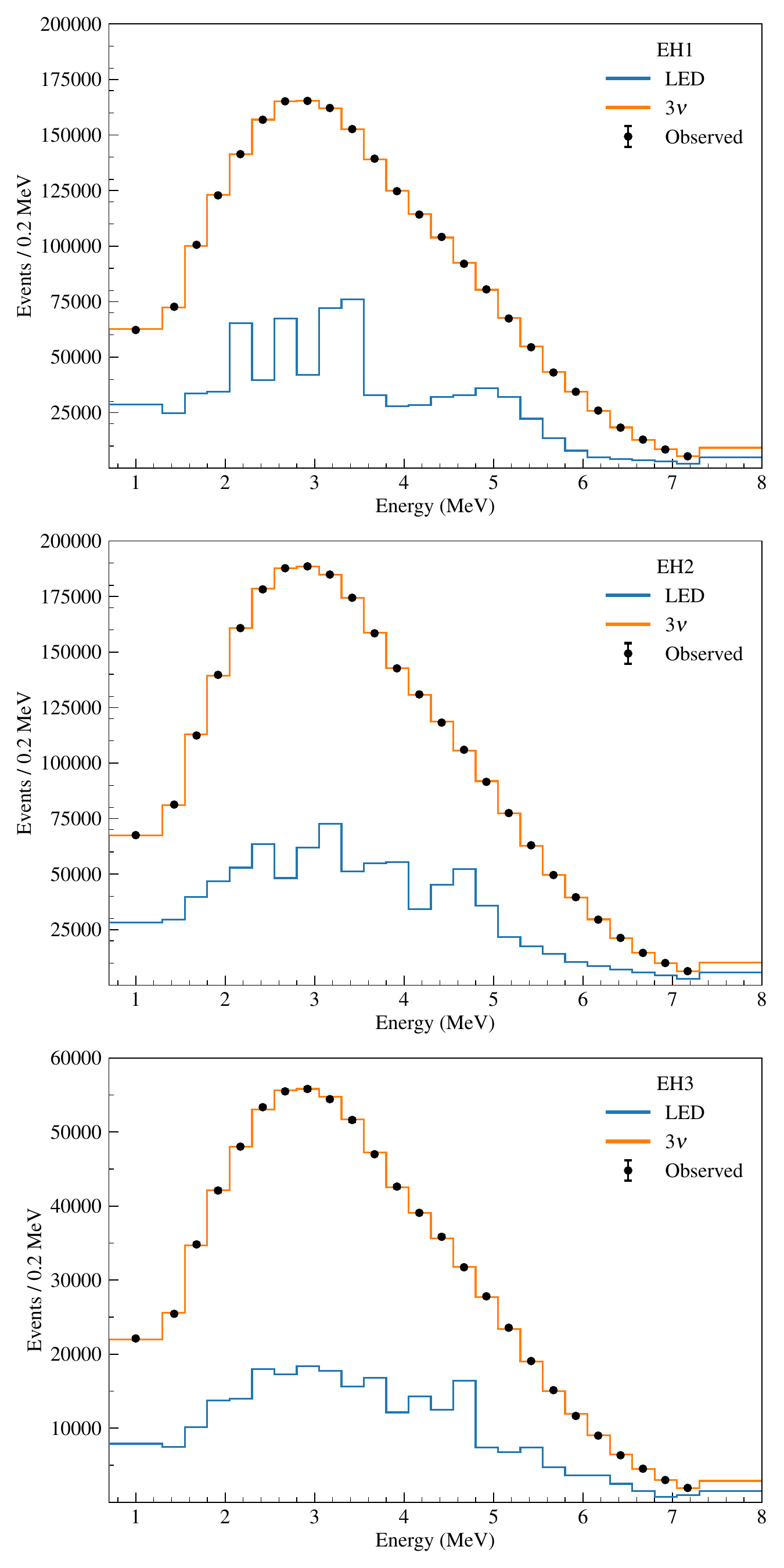}
    \caption{Observed event spectra at Daya Bay, along with the best-fit predictions for the standard 3~$\nu$ framework and the LED model. The standard oscillation parameters used are $\Delta m^2_{31} = 2.5 \times 10^{-3}~\text{eV}^2$ and $\sin^2 2\theta_{13} = 0.085$, while the LED parameters are $a = 3~\mu\text{m}$ and $m_0 = 0.1~\text{eV}$. The event distribution is shown as a function of the positron kinetic energy.}
    \label{fig:DB_best_fit}
\end{figure}

Figure~\ref{fig:DB_best_fit} shows an example of the data measured by Daya Bay over 3158 days~\cite{DayaBay:2022orm}, for the three experimental halls, along with the expected event distribution under the standard three-neutrino oscillation scenario. For this prediction, we use $\Delta m^2_{31} = 2.5\times 10^{-3}\text{eV}^2$ and $\sin^2 2\theta_{13} = 0.085$, as well as the flux parameters that best describe the data. The figure also includes the prediction assuming the presence of LED, for d=1 with $a=3\mu m$ and $m_{0} = 0.1$~eV. 

The presence of LED reduces the expected number of events due to the coupling between active and sterile states, and introduces new oscillation lengths arising from the Kaluza-Klein (KK) modes. These additional oscillations are more visible at shorter baselines, as seen in the flux measurements at EH1 and EH2.

\begin{figure*}[t]
    \centering
    \includegraphics[width=1\linewidth]{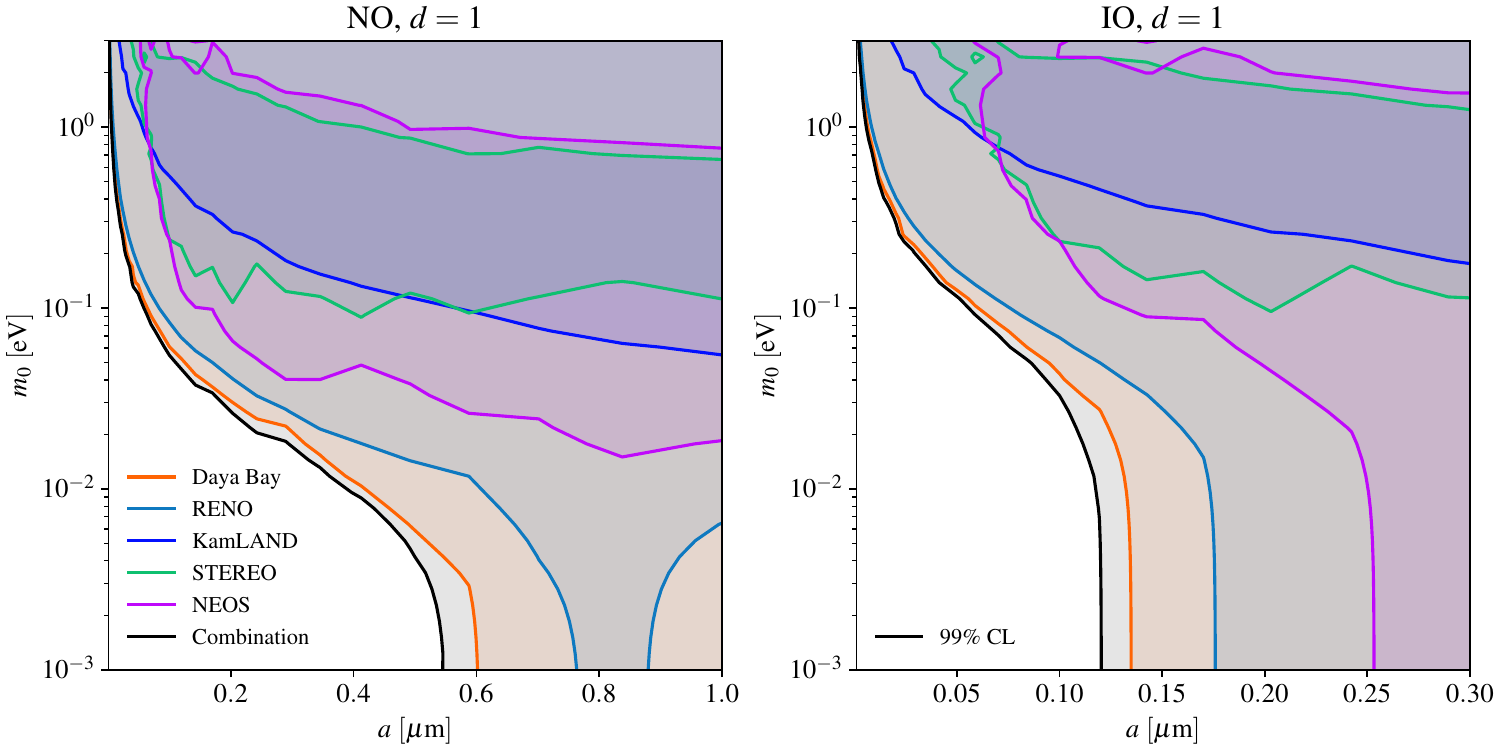}
    \caption{Large extra dimension constraints $a$ and ${\rm m}_0$ for $d=1$, i.e., 5D for the NO (left) and IO (right). We present the excluded regions from Daya Bay (orange), RENO (blue), KamLAND (dark blue), STEREO (green), NEOS (fuchsia) and their combination (black) at 99\% (full) C.L. }
    \label{fig:5D}
\end{figure*}

\section{Results}\label{sec:results} 

The combined analysis of reactor neutrino measurements across the full range of baselines where this flux has been observed allows us to search for oscillations induced by the KK-modes throughout the entire mass range accessible by the reactor spectrum. This lead us to the first global-reactor neutrino analysis aimed searching for LED.  

The results of the analysis for each experiment are shown in Fig.~\ref{fig:5D} for the five-dimensional ($d=1$) LED scenario, considering both normal ordering (left) and inverted ordering (right). The limited baseline range explored by short-baseline reactor experiments, combined with the low statistics due to the smallness of the detectors, allows these experiments to probe only a narrow region of the parameter space. Specifically, they are sensitive to masses in the range $5\times 10^{-2} \text{eV}~\leq m_{0} \leq 2\text{eV}$ for $a\geq 0.1~{\rm\mu m}$.

The region corresponding to larger neutrino masses ($m_{0}\geq 2$~eV) is excluded by measurements at longer baselines. For very light neutrinos ($m_{0}\leq 10^{-2}$~eV), medium-baseline reactor experiments ($L\sim 1~km$) provide the most stringent constraints. In the case of NO, the combined analysis indicates that the radius of the extra dimension must be smaller than $a \leq 0.58~{\rm\mu m}$ at 99\% CL. For IO, where the impact of LED on neutrino evolution is more pronounced, the bound becomes more stringent $a \leq 0.12~{\rm\mu m}$ at 99\% CL. 

\begin{figure*}
    \centering
    \includegraphics[width=1\linewidth]{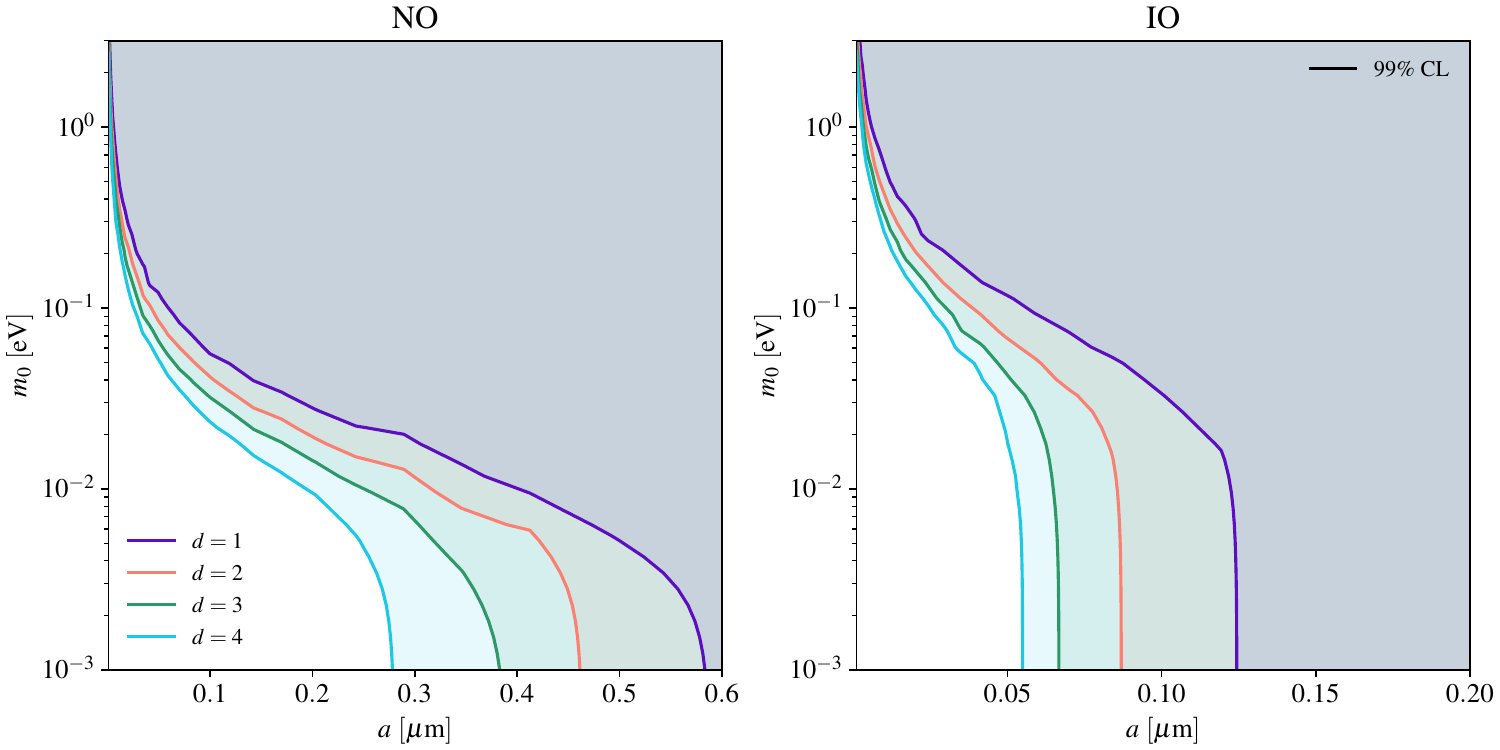}
    \caption{Large extra dimension constraints $a$ and ${\rm m}_0$ for the NO (left) and IO (right). We present the excluded regions for $d=1$ (purple), $d=2$ (orange), $d=3$ (green) and $d=4$ (turquoise) at 99\% C.L.}
    \label{fig:LED-RENO+DB_IO}
\end{figure*}

When considering scenarios with a larger number of extra dimensions, $d>1$, the amplitude of the new oscillation modes becomes more pronounced, leading to a stronger impact on neutrino evolution. As a result, the constraints on the radius of the extra dimension become tighter for both normal and inverted mass orderings, as illustrated in Figure~\ref{fig:LED-RENO+DB_IO}. In these cases, we have focused on experiments with baselines near the kilometer scale, Daya Bay and RENO, since, as observed in the five-dimensional scenario, the sensitivity is dominated by those measurements. 

For normal ordering and light neutrino masses, ($m_{0}\leq 10^{-2}$~eV), the bounds on the radius of the extra dimension become increasingly stringent as the number of dimensions increases. The constraints are: $a\leq 0.45~{\rm\mu m}$ for $d=2$, $a\leq 0.38~{\rm\mu m}$ for $d=3$ and $a\leq 0.28~{\rm\mu m}$ for $d=4$\footnote{As noted before, we note that the case of $d=4$ might be more heavily dependent on the UV completion. However, we derived constraints without assuming any specific case.}. As in the case with a single extra dimension, the impact of LED on neutrino evolution is more significant under inverted ordering, resulting in even stronger bounds: $a\leq 0.09~{\rm\mu m}$ for $d=2$, $a\leq 0.06~{\rm\mu m}$ for $d=3$ and $a\leq 0.055~{\rm\mu m}$ for $d=4$.

\section{Conclusions}\label{sec:concl} 

The possible existence of additional spatial dimensions beyond the three observed ones has profound implications for particle physics. 
In particular, their presence could address some long-standing open problems, such as the hierarchy problem and the smallness of neutrino masses. 
In the latter case, the introduction of additional right-handed neutrinos that are allowed to propagate through the extra dimensions gives rise to Kaluza-Klein modes that mix with the standard neutrinos, thereby potentially modifying their oscillations.

In this work, we have updated the constraints on the size of the extra dimensions using the complete data sets from several reactor antineutrino experiments, namely Daya Bay, RENO, KamLAND, STEREO, and NEOS. 
We find that the bound is dominated by Daya Bay data, with only a marginal contribution from RENO. 
Our results show that the limits improve by approximately $\sim 20\%$ ($\sim 25\%$) for normal (inverted) ordering in the limit where the lightest neutrino is massless. 

Additionally, by numerically computing the oscillation probability for different values of $d$, the number of extra dimensions, we have derived for the first time constraints on $d = 2, 3, 4$ using reactor data under the assumption that all extra dimensions share the same radius. 
Since the oscillation probabilities deviate significantly from the standard case as the number of extra dimensions increases, due to the additional mixing induced by the KK modes, we find that the bounds become more stringent for larger $d$. 
Specifically, for $d = 2~(4)$, the limit on the size of the extra dimensions is $a \lesssim 0.45~{\rm \mu m}~(0.28~{\rm \mu m})$ for NO at the $99\%$ confidence level. 
For IO, the corresponding bounds are $a \lesssim 0.09~{\rm \mu m}~(0.07~{\rm \mu m})$ for $d = 2~(4)$ at the same confidence level. 
These results highlight the constraining power of neutrino oscillation data in exploring scenarios beyond the SM. 
The next generation of neutrino oscillation experiments will further enhance our ability to probe such physics.

\begin{acknowledgments}

Y.F.P.G. has been supported by the Consolidaci\'on Investigadora grant CNS2023-144536 from the Spanish Ministerio de Ciencia e Innovaci\'on (MCIN) and by the Spanish Research Agency (Agencia Estatal de Investigaci\'on) through the grant IFT Centro de Excelencia Severo Ochoa No CEX2020-001007-S.  IMS is supported by STFC grant ST/T001011/1.

\end{acknowledgments}

\bibliographystyle{apsrev4-1}
\bibliography{LEDreactors}
\appendix

\section{Simulation Details}\label{sec:simdet}

In this section, we provided a detailed description of the data simulated included in this analysis. 

\begin{figure*}
    \centering
    \includegraphics[width=0.48\linewidth]{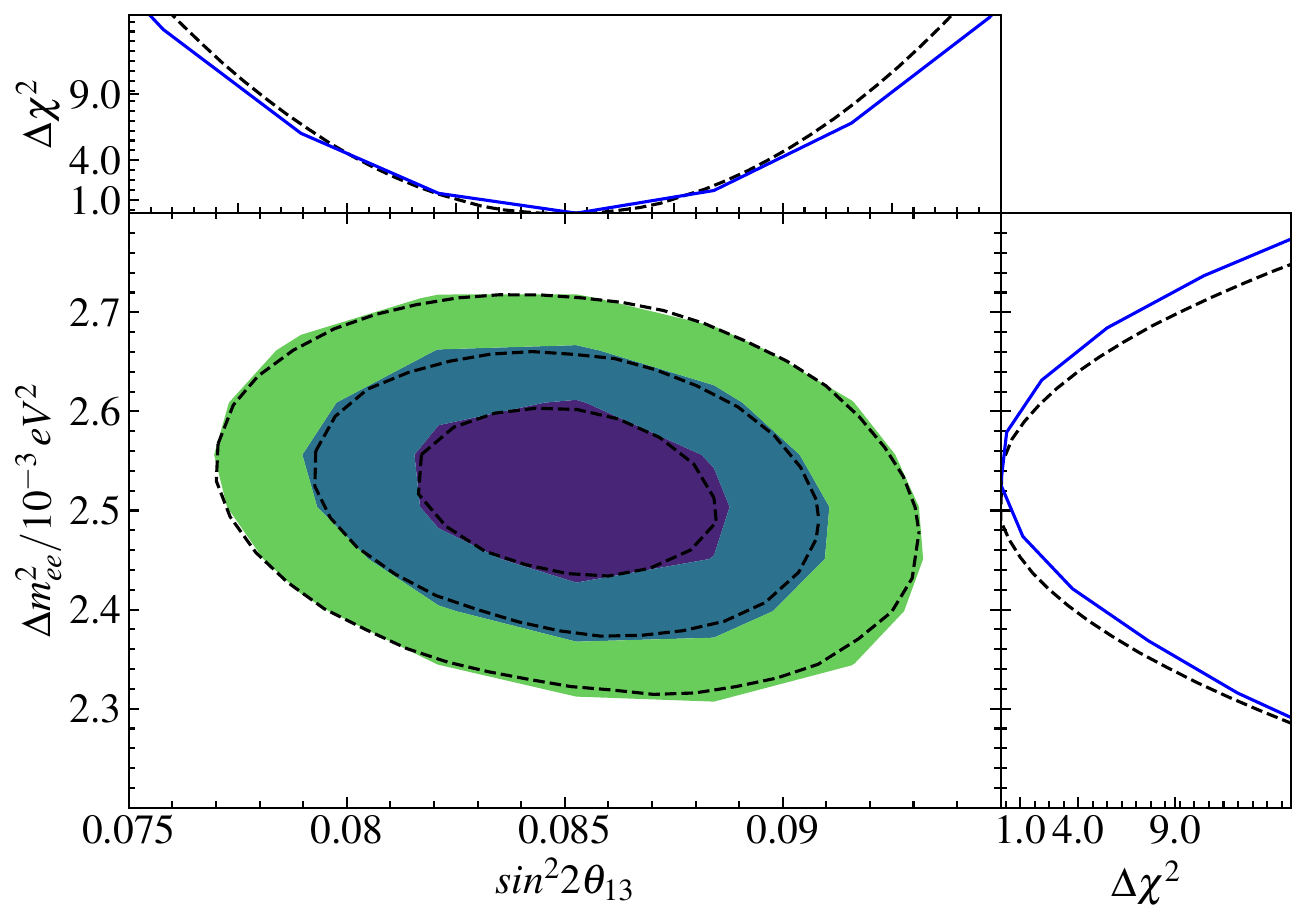}
    \includegraphics[width=0.48\linewidth]{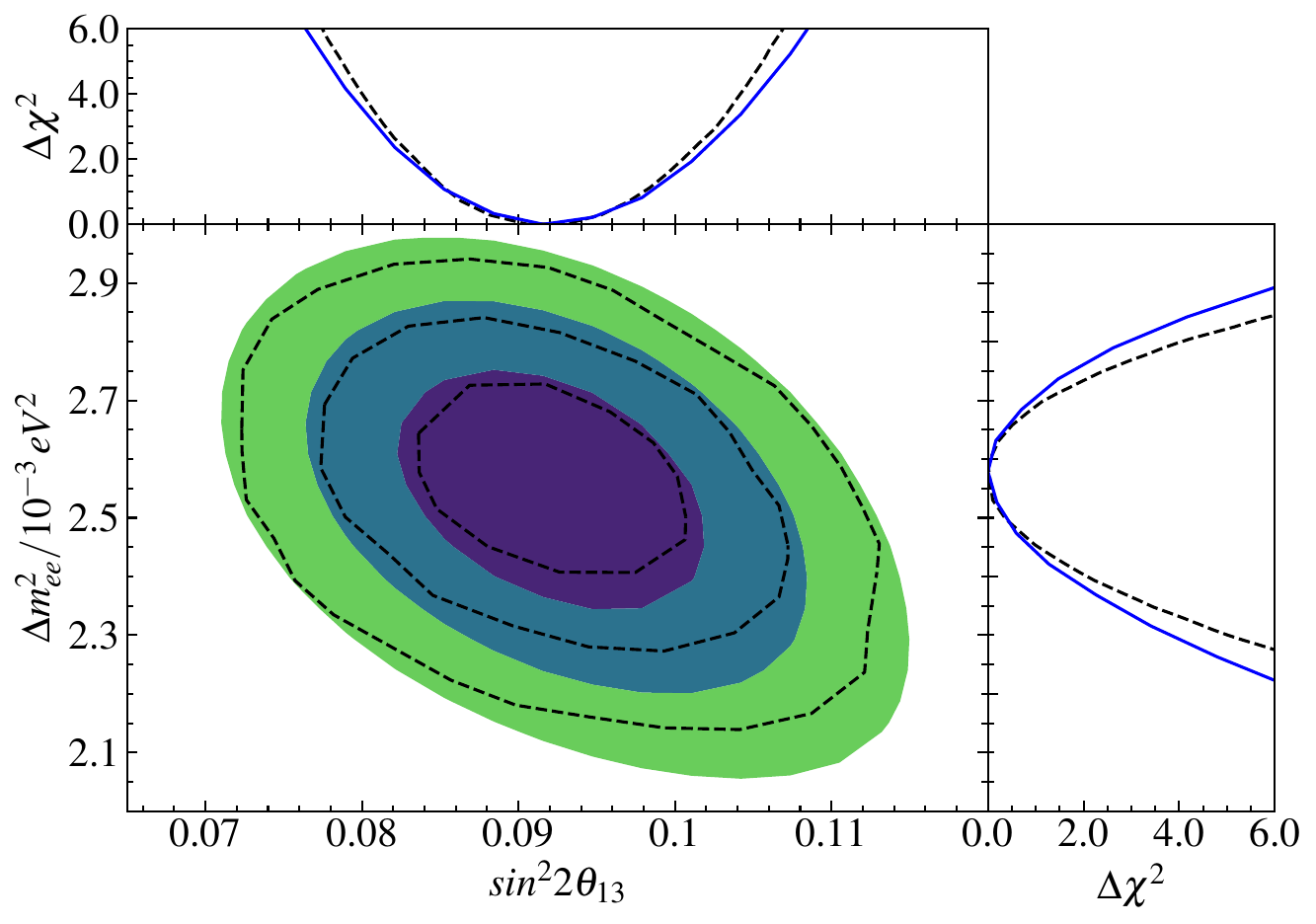}
    \caption{ Analysis of Daya Bay (left) and RENO (right) final data release considering the $3\nu$ scenario. The results of this analysis are given by the colored regions, where each region corresponds to the $1\sigma,2\sigma$ and $3\sigma$. See the text for details about the analysis. The results are compared with Daya Bay~\cite{DayaBay:2022orm} and RENO~\cite{RENO:2024msr} results shown by the dashed lines.}
    \label{fig:RENO+DB-3nu}
\end{figure*}

\subsection{Daya Bay}

The final analysis of the Daya Bay experiment~\cite{DayaBay:2022orm} incorporates data collected over 3,158 days of operation at the Daya Bay–Ling Ao nuclear power facility in Shenzhen. The experiment comprised six 2.9-GW$_\text{th}$ reactors, distributed across three underground experimental halls designated EH1, EH2, and EH3. Up to eight antineutrino detectors (ADs) were deployed to capture the emitted $\bar{\nu}_e$ flux. Each AD contained approximately 20 tonnes of liquid scintillator doped with 0.1\% gadolinium by weight (GdLS), used to detect inverse beta decay (IBD) events~\cite{Yeh:2007zz, Beriguete:2014gua, DayaBay:2015kir, DayaBay:2014cmr}.

The flux at each detector hall is calculated as the sum of contributions from all reactors to all detectors within that hall:
\begin{align}
\Phi^{p, \text{EH}}_{\bar{\nu}_e} = \sum_{r} \sum_{d} 
N_p \,M_{d} \, \varepsilon_{d} \, W_{\mathrm{th}}^{(r)} \,
P_{\bar{\nu}_e \rightarrow\bar{\nu}_e}
 \frac{S_\nu(E_\nu)}{4\pi L_{d}^{(r)\, 2}},
\end{align}
where $N_{p}$ is the number of free protons, given by $N_p = N_{A} {\rm f_H I_H m_H}$, with $N_{A}$ being the Avogadro's number, ${\rm f_H}$ the hydrogen fraction, ${\rm I_H}$ the isotope abundance, and ${\rm m_H}$ the atomic mass of Hydrogen. The number of targets is weighted by the detector efficiency ($\varepsilon_d$). 

The anti-electron neutrino flux depends on the thermal power of the r-th detector, $W_{\mathrm{th}}^{(r)}$, and the anti-neutrino spectra ($S_\nu(E_\nu)$), given by Daya Bay masurements~\cite{DayaBay:2019yxq}. The experiment run in three configuration periods using six (6AD), eight (8AD) and seven (7AD) detectors. Reactor power data were only provided for the 7AD period; for the 6AD and 8AD periods, we used averaged values.

The flux of $\bar{\nu}_e$ is further weighted by the survival probability $P_{\bar{\nu}_e \rightarrow \bar{\nu}_e}$. To obtain the expected number of events, the antineutrino flux is convolved with the inverse beta decay cross-section~\cite{Vogel:1999zy} and the duration of each data-taking period: 217 days for the 6AD configuration, 1,524 days for 8AD, and 1,417 days for 7AD.

In this analysis, we fit the observed data using a Gaussian $\chi^2$ function:
\begin{align}
\chi^2 &= \sum_{b,\, \mathrm{EH},\, p} \frac{\left[(1+\alpha^{\mathrm{norm}}_{\mathrm{EH}})\left(\frac{E_b}{E_{\mathrm{piv}}}\right)^{\gamma_{\mathrm{EH}}} N^{\mathrm{exp}}_{b\mathrm{EH}p} - N^{\mathrm{obs}}_{b\mathrm{EH}p}\right]^2}{N^{\mathrm{obs}}_{b\mathrm{EH}p}} \nonumber \\
&\quad + \sum_{\mathrm{EH}} \left[ \left(\frac{\alpha^{\mathrm{norm}}_{\mathrm{EH}}}{\sigma_{\mathrm{norm}}}\right)^2 + \left(\frac{\alpha^{\mathrm{bkg}}_{\mathrm{EH}}}{\sigma_{\mathrm{bkg}}}\right)^2 + \left(\frac{\gamma_{\mathrm{EH}}}{\sigma_{\gamma}}\right)^2 \right]
\label{eq:DB-chi2}
\end{align}
We sum the contributions to the $\chi^2$ from each energy bin ($b$), each data-taking period ($p$), and each experimental hall (EH). The expected number of events in each bin is given by the sum of the expected number of electron antineutrinos arriving at the detector and the background contribution:
\begin{equation}
N^{\mathrm{exp}}_{b\mathrm{EH}p} = N^{\mathrm{IBD}}_{b\mathrm{EH}p}(\theta_{13}, \Delta m^2_{ee}) + (1 + \alpha^{\mathrm{bkg}}_{\mathrm{EH}}) N^{\mathrm{bkg}}_{b\mathrm{EH}p},
\end{equation}
where $N^{\mathrm{bkg}}_{b\mathrm{EH}p}$ is the total number of expected background events. In this analysis, we assume a flux normalization uncertainty $\alpha^{\mathrm{norm}}_{\mathrm{EH}}$ with $\sigma_{\mathrm{norm}} = 0.22\%$. We also include a background normalization uncertainty of $\sigma_{\mathrm{bkg}} = 10\%$. The dominant uncertainty arises from a modification of the energy spectrum of the flux, parameterized by a tilt factor $(E/E_{\mathrm{piv}})^{\gamma}$, with $E_{\mathrm{piv}} = 2.5\,\mathrm{MeV}$ and $\sigma_{\gamma} = 1\%$. Figure~\ref{fig:RENO+DB-3nu} shows a comparison between the results obtained in this analysis (regions) and Daya Bay collaboration (dashed lines) in the standard $3\nu$ scenario.

\subsection{RENO}

The RENO experiment measured the electron antineutrino flux from the Hanbit nuclear power plant in Yeonggwang. The final data analysis includes 3,800 days of data collection. The experiment consists of two identical detectors: the near detector is located 294 meters from the center of the six 2.8-GW$_\text{th}$ reactors, while the far detector is located 1,383~m away. Details of the detector design and data acquisition system can be found in Refs.~\cite{RENO:2016ujo, RENO:2010vlj}.

Similar to the Daya Bay analysis, we performed a simulation to reproduce the number of observed events. The simulation follows the same methodology as used for Daya Bay. The chi-squared function used in the RENO analysis takes the following form:
\begin{align}
\chi^2 &= \sum_{N,F} \sum_{b} \frac{[(E_{b}/E_{\rm piv})^{\gamma}N^{\rm exp}_{b}  -  N^{\rm obs}_{b}]^2}{N^{\rm obs}_{b}} \nonumber \\
&+\sum_{N,F}\Bigg[ +\left(\frac{\alpha^{\rm bkg}_{N,F}}{\sigma_{\rm bkg}^{N, F}}\right)^2 + \left(\frac{\alpha^{\rm Cf}}{\sigma_{\rm Cf}}\right)^2\nonumber + \left(\frac{\alpha^{\rm LiHe}}{\sigma_{\rm LiHe}}\right)^2  \\
&+\left(\frac{\gamma}{\sigma_{\gamma}}\right)^2\Bigg] + \sum_{r, N,F}\left(\frac{\alpha^{\rm flux}_r}{\sigma_{\rm flux}}\right)^2.
\end{align}
We sum the contributions to the $\chi^2$ from both the near and far detectors and across all energy bins. We include uncorrelated uncertainties for each background component, such as contamination from $^{252}\mathrm{Cf}$ ($\alpha^{\rm Cf}$) and the isotopes $^9\mathrm{Li}$ and $^8\mathrm{He}$ ($\alpha^{\rm LiHe}$), which are produced inside the detector by cosmic muons. Background predictions are taken from Ref.~\cite{RENO:2024msr}. For these backgrounds, we assume a common uncertainty of $\sigma_{\rm Cf, LiHe} = 4.15\%$.

We also account for variations in the energy dependence of the neutrino spectrum using a tilt factor, with an associated uncertainty of $\sigma_\gamma = 0.4\%$. Additionally, we include uncertainties in the contributions from each isotope and the relative flux between the near and far detectors, all parameterized by $\alpha^{\rm flux}_r$, with an uncertainty of $\sigma_{\rm flux} = 0.2\%$. The result of the analysis in the case of the $3\nu$ scenario are compared to the RENO results in Fig~\ref{fig:RENO+DB-3nu} (right.)

\subsection{NEOS}

The NEOS detector, like the RENO detectors, is located at the Hanbit nuclear power plant. Specifically, it is located at $23.7\pm 0.3$~m from the fifth reactor core. The NEOS data analysis includes 180 days of data acquisition. Further details about the experiment can be found in Refs.~\cite{NEOS:2016wee, RENO:2020hva}.

The primary goal of NEOS is to search for sterile neutrinos. To achieve this, the experiment compares the event distribution observed in NEOS with that measured by the RENO near detector. We follow the methodology described in Ref.~\cite{Dentler:2017tkw} to estimate the NEOS sensitivity.  The chi-squared function used in the analysis is defined as:
\begin{equation}
\chi^2 = \sum_{i,j} \bigg(O^i - P^{i}\bigg) V^{-1}_{ij}
\bigg(O^j - P^{j}\bigg),
\end{equation}
where $O^i$ is the ratio between the observed spectrum in NEOS and the prediction from RENO, and $P^{i}$ is the ratio between the expected event distribution in NEOS under the LED hypothesis and the distribution expected in the standard $3\nu$ scenario. To estimate the expected number of events in NEOS, we use the predicted event distribution from RENO, which includes the neutrino evolution to the RENO near detector.

\subsection{Stereo}

The STEREO experiment measures the electron antineutrino flux from the RHF (Réacteur à Haut Flux) nuclear reactor, located at the ILL research center in Grenoble, France~\cite{STEREO:2019ztb}. The antineutrino spectrum is measured using six identical detector cells positioned between 9.4~m and 11.2~m from the reactor core. The primary goal of the experiment is to search for sterile neutrinos with masses around the eV scale.

The reactor, moderated by heavy water, operates at a nominal thermal power of 58.3~MW, although in practice the power is typically maintained between 50~MW and 56~MW. The detector uses a 1.6-ton gadolinium-loaded liquid scintillator target to detect antineutrinos produced via IBD.

To estimate the detector's sensitivity to large extra dimensions (LED), we follow the analysis performed in the sterile neutrino search~\cite{hepdata.92323}. The $\chi^2$ function used is defined as:
\begin{align}
\chi^2&=\sum_{l=1}^{N_{\mathrm{cells}}}\sum_{i=1}^{N_{\mathrm{Ebins}}}\left(\frac{A_{l,i}-\phi_i M_{l,i}}{\sigma_{l,i}}\right)^2
    +\sum_{l=1}^{N_{\mathrm{cells}}}\left(\frac{\alpha_l^{\mathrm{EscaleU}}}{\sigma_l^{\mathrm{EscaleU}}}\right)^2 \nonumber
    \\&+\left(\frac{\alpha^{\mathrm{EscaleC}}}{\sigma^{\mathrm{EscaleC}}}\right)^2+\sum_{l=1}^{N_{\mathrm{cells}}}\left(\frac{\alpha_l^{\mathrm{NormU}}}{\sigma_l^{\mathrm{NormU}}}\right)^2, \label{eq:stereo-chi2}
\end{align}
where the sum runs over all detector cells ($l$) and energy bins ($i$). Here, $A_{l,i}$ denotes the measured IBD rates, and $M_{l,i}$ represents the expected IBD rates, which depend on both the oscillation parameters and the nuisance parameters $\vec{\alpha}$. The $\phi_i$ values are free normalization parameters for each energy bin $i$.

The parameters $\alpha^{\mathrm{EscaleC}}$ and $\alpha^{\mathrm{EscaleU}}$ account for the correlated and uncorrelated energy scale uncertainties of the detectors, respectively. The nuisance parameters $\alpha^{\mathrm{NormU}}$ represent cell-to-cell uncorrelated normalization uncertainties. The statistical uncertainties in Eq.~\ref{eq:stereo-chi2} are given by $\sigma_{l,i} = \sigma_{l,i}(\phi_i M_{l,i})$, which depend on the predicted event rate.

\subsection{KamLAND}

KamLAND measured the electron antineutrino flux from nuclear power plants across Japan between 2002 and 2012~\cite{KamLAND:2013rgu}. Over 2,991 days of data collection, KamLAND observed neutrinos with an average baseline of approximately 180~km, divided into three distinct data-taking periods. The detector consists of 1~kton (LS). The primary goal of KamLAND was to probe long-baseline neutrino oscillations, and to date, it has provided the most precise determination of the mass-squared difference $\Delta m^2_{21}$. As shown in several previous analysis~\cite{Machado:2011kt}, KamLAND also constrains the high-mass region of the LED parameter space.

In this analysis, we reproduce the number of events observed by KamLAND ($T$) across the three data-taking periods. Taking into account the distribution of reactors throughout Japan and their contributions to the total flux, we estimate the $\overline{\nu_{e}}$ flux at the detector site in the presence of LED effects. We also include the main sources of background, which consist of: geoneutrinos produced by the decay of radioactive elements such as uranium (U) and thorium (Th) in the Earth, neutrons generated by the interaction of alpha particles from radioactive polonium ($^{210}$Po) with carbon ($^{13}$C) via the reaction \({}^{13}C(\alpha ,n)^{16}O\), and accidental backgrounds. The analysis follows a Poissonian log-likelihood defined as:
 \begin{align}
      \chi^2 &= \sum_{i} T(a,m_{0})_{i} - O_{i} + O_{i} \log\left(\frac{O_{i}}{T(a,m_{0})_{i}}\right)\nonumber \\
      &+\sum^{bkg}_{b} \left(\frac{\alpha_{b} - 1}{\sigma_{b}}\right)^2
 \end{align}
where $O$ is the observed number of events in energy bin i, and the sum runs over all energy bins and the three data-taking periods. In this analysis, we assume a 2\% Gaussian uncertainty in the energy reconstruction. The final term accounts for uncertainties in the background contributions. We consider a 5\% uncertainty for the accidental background and a 10\% uncertainty for the \({}^{13}C(\alpha ,n)^{16}O\) background, while the remaining background components are left unconstrained.

\end{document}